\documentclass[conference]{IEEEtran}

\PassOptionsToPackage{
	pdfauthor={Martin Geier, Dominik Faller, Marian Braendle and Samarjit Chakraborty},
	pdftitle={Cost-effective Energy Monitoring of a Zynq-based Real-time System including dual Gigabit Ethernet},
	pdfsubject={Cost-effective Energy Monitoring of a Zynq-based Real-time System including dual Gigabit Ethernet (Extended arXiv version)},
}{hyperref}

\usepackage[utf8]{inputenc}
\usepackage{amsmath,amssymb}
\usepackage{listings}
\usepackage{url}
\usepackage{graphicx}
\usepackage[hidelinks=true]{hyperref}

\begin{document}

\title{Cost-effective Energy Monitoring of a Zynq-based Real-time System including dual Gigabit Ethernet}
\author{\IEEEauthorblockN{\href{https://orcid.org/0000-0003-2481-9873}{Martin Geier\,\includegraphics[height=\fontcharht\font`\O]{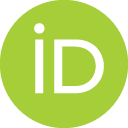}}, Dominik Faller, Marian Brändle and Samarjit Chakraborty}
\IEEEauthorblockA{Chair of Real-Time Computer Systems\\
Technical University of Munich\\
\href{mailto:geier@rcs.ei.tum.de}{geier}/\href{mailto:faller@rcs.ei.tum.de}{faller}/\href{mailto:braendle@rcs.ei.tum.de}{braendle}/\href{mailto:chakraborty@rcs.ei.tum.de}{chakraborty}@rcs.ei.tum.de}
}
\maketitle

\begin{abstract}
The ongoing integration of fine-grained power management features already established in CPU-driven Systems-on-Chip (SoCs) enables both traditional Field Programmable Gate Arrays (FPGAs) and, more recently, hybrid Programmable SoCs (pSoCs) to reach more energy-sensitive application domains (such as, e.g., automotive and robotics).
By combining a fixed-function multi-core SoC with flexible, configurable FPGA fabric, the latter can be used to realize heterogeneous Real-time Systems (RTSs) commonly implementing complex application-specific architectures with high computation and communication (I/O) densities.
Their dynamic changes in workload, currently active power sav-\linebreak{}ing features and thus power consumption require precise voltage and current sensing on all relevant supply rails to enable dependable evaluation of the various power management techniques.

\noindent
In this paper, we propose a low-cost 18-channel 16-bit-resolution measurement (sub-)system capable of 200~kSPS (kilo-samples per second) for instrumentation of current pSoC development boards.
To this end, we join simultaneously sampling an\-a\-log-to-dig\-i\-tal converters (ADCs) and analog voltage/current sensing circuitry with a Cortex M7 microcontroller using an SD card for storage.
In addition, we propose to include crucial I/O components such as Ethernet PHYs into the power monitoring to gain a holistic view on the RTS's temporal behavior covering not only computation on FPGA and CPUs, but also communication in terms of, e.g., reception of sensor values and transmission of actuation signals.
We present an FMC-sized implementation of our measurement system combined with two Gigabit Ethernet PHYs and one HDMI input.
Paired with Xilinx' ZC702 development board, we are able to synchronously acquire power traces of a Zynq pSoC and the two PHYs precise enough to identify individual Ethernet frames.
\end{abstract}

\IEEEpeerreviewmaketitle

\section{Introduction}
\enlargethispage{0.1575mm}
\noindent
Although continuous advances in process technology and design automation tools sustain the development of increasingly complex Systems-on-Chip (SoCs), the resulting computational and communicational densities also elevate the energy demand of such processing platforms.
This not only holds true for SoCs found in everyday devices (e.g., smartphones), but also for those operating under predefined latency constraints in, e.g., safety-critical environments~\cite{7746761}.
With such Real-time Systems (RTSs) pervading more supply- and cooling-constrained application domains such as automotive and robotics, power saving capabilities (and overlying management techniques) have become an additional design goal.
Although traditionally lagging behind in this regard, current Programmable SoCs (pSoCs) that combine fixed-function multi-core SoCs with configurable FPGA fabric thus have started to integrate fine-grained power management features~\cite{5272538}.
This includes not only well-known methods from Application-Specific Integrated Circuit (ASIC) design (such as glitch-free clock gates and multiplexers), but also architecture-specific device features (such as automated power gating of uninstantiated memory resources~\cite{xilinx-ug473}).
Combining various primitives for both clock generation (e.g., clock management tiles) and distribution (e.g., global, regional and I/O clock trees)~\cite{xilinx-ug472} of the underlying FPGA architectures with their capabilities for Dynamic Partial Reconfiguration (DPR) enables energy-aware runtime strategies on current pSoCs~\cite{7857177}.

As the evaluation of (existing and) novel methods requires extensive performance figures in terms of energy consumption and, e.g., throughput, simulation-, model- and measurement-based approaches are widely used~\cite{4068926,1568543}.
Whilst the former rely on either coarse activity rates (e.g., Xilinx Power Estimator) or more accurate simulation-driven waveforms, the latter involve actual voltage and current readings from a physical device-under-test (DUT).
The soundness reached by such measurements heavily depends on used approach and equipment, which range from external, supply-side oscilloscope readings to per-rail traces captured by either external meters or on-board monitors -- all with distinct limits in resolution and accuracy.

In this paper, we propose a cost-efficient energy monitoring (sub-)system for instrumentation of current pSoC development boards.
Three six-channel, simultaneously sampling analog-to-digital converters (ADCs) with the necessary analog circuitry for voltage and current sensing enable the proposed system to capture nine or more supply rails by either remote instrumentation (via V/I probes routed to suitable pads of a development board) or direct integration (on the same card).
Driven by a Cortex M7 microcontroller, the resulting system is capable of synchronously capturing its 18 analog channels, a trigger line and, if available, measurements from external PMBus power controllers to an SD card.
In addition, we propose to extend the power monitoring to those I/O components (such as Ethernet PHYs) interfacing the RTS to its outside world, which results in a holistic view on the overall, usually latency-constrained\linebreak temporal behavior covering reception of sensor values, the processing within the pSoC to transmission of actuation signals.

\enlargethispage{0.1575mm}
We present an implementation of the proposed measurement system on an FPGA Mezzanine Card (FMC) that also includes and monitors two Gigabit Ethernet PHYs and an HDMI input.
Achieving sampling rates of beyond 200~kSPS (kilo-samples per second) and current resolutions down to tens of $\mu$A, our proposed system implements an intermediate solution between high-resolution, but single-channel external oscilloscopes and multi-rail, but low-resolution on-board monitors.
Combining a ZC702 development board with our FMC prototype (featuring monitored I/O components), we are able to identify individual Ethernet frames and processing phases of a high-speed Visual Servoing application scenario in a cost-effective manner.

\section{Related Work}
\noindent
Whilst static and dynamic energy optimization techniques for both ASICs and FPGAs are a well-explored field, our extensive literature search revealed that most work relies either on power modeling or (in case of FPGAs and recently pSoCs) relatively low-resolution measurements.
\emph{Modeling} not only requires in-depth knowledge of the various underlying design and process technologies, but also some notion of circuit activity to characterize the overall energy consumption~\cite{4068926}.
Whilst process parameters such as CMOS transistor sizes, voltage levels and temperature primarily influence a device's static power (due to mostly fixed leakage currents), the design's clock frequencies, signal activities and net fan-out mainly affect its dynamic power consumption.
Thus, apart from accurate ASIC/process information (commonly integrated/hidden in the various EDA tools), dependable power modeling requires reliable estimates of each node's switching activity.
For FPGA- and pSoC-based designs, available tools range from spreadsheets -- whose accuracy heavily depends on reliable, user-supplied usage figures per function block (such as logic, memories, clocking and I/O ports) -- to cascaded simulation/modeling toolchains utilizing activity waveforms gathered from functional simulation~\cite{7529097}.

\emph{Measurement}-based approaches, on the other hand, are also challenging for a number of reasons.
First, most FPGAs rely on multiple independent supply rails dedicated to their various internals such as Configurable Logic Blocks (CLBs), Block-RAMs (BRAMs), Phase-Locked Loops (PLLs) and I/O banks.
The (hardcore) CPUs, memory controllers and other function blocks of current pSoCs commonly require a similar number of (additional) supplies, typically resulting in around ten rails on a pSoC board.
For comprehensive power monitoring suitable for a multitude of designs (each with its unique distribution of demand across all rails), thus both voltage and current of several rails have to be captured simultaneously.
As current measurement requires either a shunt resistor or a (more costly) hall sensor on the on-board trace or, if electrically feasible, an off-board wire, multi-rail coverage is mostly implemented with individual shunt resistors on the Printed Circuit Board (PCB).
Whilst relatively easy to add to (hard to design) custom PCBs, many development boards already come with such resistors as part of their integrated supply solution.
In case no shunts are available on the PCB, only the overall power consumption can be measured on the supply side~\cite{5694272,6822344,7857177} or, as an extreme case, a rail might be disconnected and driven externally~\cite{1568543}.
\\
On boards with appropriate shunt resistors on the supply rails of interest, the resulting sense (and actual rail) voltages can be captured either off-board or internally.
External solutions range from standalone oscilloscopes via USB analog/logic analyzers to fully-fledged data acquisition (DAQ) solutions (such as National Instruments's PXI series).
Traditional oscilloscopes are not only available with a wide choice of measurement parameters (e.g., bandwidth, resolution, accuracy, number of channels\linebreak and analysis features such as storage to USB), various vendors and resulting cost, but also offer relatively simple, standalone operation.
USB-based analyzers and most DAQ solutions, in contrast, require host PCs for control and data acquisition that, depending on their memory sizes, limit the maximum capture lengths.
Whilst the former tend to have limited resolution (e.g., 12~bits in case of a Saleae Logic Pro 8), the latter offer a wide variety of measurement boards easily, e.g., reaching 24~bits at 200~kHz on two channels (PXI-4461) or even higher sampling rates of 4~MHz at 16~bits on four channels (PXIe-6124).
One key difference between the above solutions is the cost incurred for hardware and sometimes required software licenses.
Whilst oscilloscopes and USB analyzers start at hundreds of dollars, a single DAQ card alone costs ten times more.
Due to chassis, PCIe controller and software licenses, the cost of a moderately equipped DAQ system easily exceeds tens of thousands.
\\
Internal solutions, however, capture the (usually preamplified) current sense and (actual) rail voltages using on-board ADCs.
The latter are commonly integrated in either the system-level supply controllers (e.g., TI's UCD92xx series) or every sense voltage instrumentation amplifier (such as TI's INA231)~\cite{6853134,7238104}.
Once digitized, the rail's voltage and current readings are available via an $\text{I}^{2}\text{C}$/PMBus interface.
Although those devices are comparatively cheap (in the single-digits dollars), they both not only require an external master device for data retrieval and storage, but also have limited sampling rates due to their serial interface.
In (the common) case all devices share a single bus, the overall sampling rate is further reduced due to sequential readout (to, e.g., a few hundred SPS)~\cite{6853134}.
Although dedicated current monitors reach resolutions of a few hundred~$\mu$A with typical shunt resistors, system-level controllers tend to be less accurate due to wider value ranges and limited PMBus number representations.
In case of TI's UCD9248 used in various Zynq designs, the current resolution is in the order of tens of mA.

\begin{figure}
	\centering
	\includegraphics[width=\linewidth]{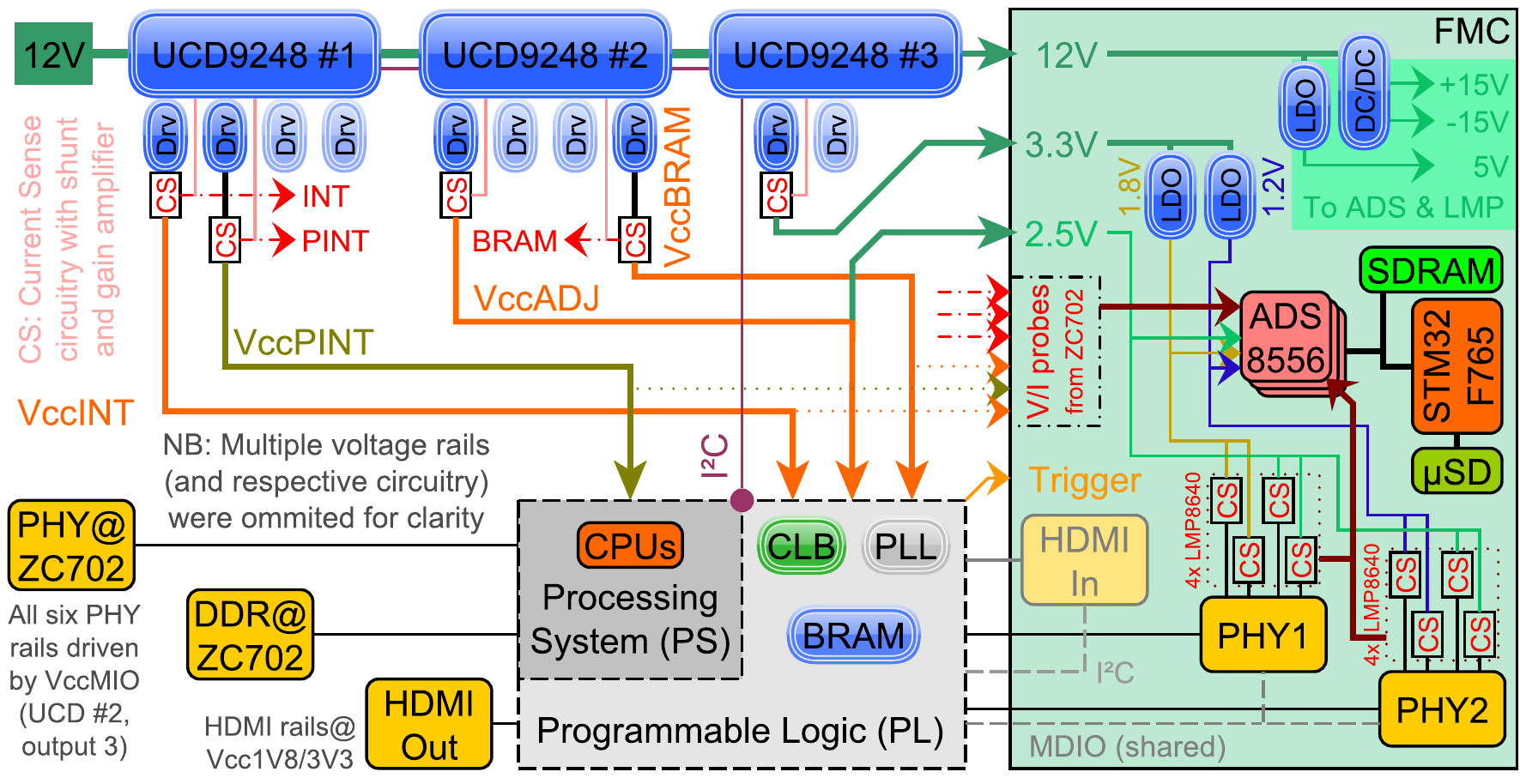}
	\vspace{-6mm}
	\caption{Internals of ZC702 Development Board and our FMC Prototype (right)}
	\label{fig1}
	\vspace{-6mm}
\end{figure}

In comparison to above measurement approaches, the proposed system implements an intermediate design choice w.r.t. several parameters.
Resolution-wise, our solution ranks above common system-level supply controllers and USB analyzers, but below oscilloscopes and DAQ systems.
In terms of sampling rate, our system (reaching over 200~kSPS) exceeds $\text{I}^{2}\text{C}$-based solutions by three orders of magnitude, although oscilloscopes and (selected) high-speed DAQ cards reach hundreds of MHz.
Whilst the maximum capture length of oscilloscopes is limited by internal memory, most non-standalone solutions can, to a certain degree/cost, be equipped with additional RAM (USB analyzers, DAQ systems) or flash storage (our solution).

It also should be noted that, unsurprisingly, serious discrepancies between modeling and measurement have already been observed due to their various influencing factors~\cite{4068926,6822344,7238104}.
Nevertheless, with each method having its strengths and weaknesses, both remain valuable tools for energy optimization.

\section{Proposed Measurement System Architecture for Energy Monitoring of pSoCs and I/O Devices}
\noindent
During the concept of the proposed measurement system, we faced the following key design questions.
\emph{Measurement data retrieval and storage} could be mapped to either the DUT (i.e., Zynq pSoC) itself or a dedicated microcontroller unit (MCU).
Although the former is in line with existing internal solutions (in which the DUT itself queries the power monitors via $\text{I}^{2}\text{C}$), it also skews the DUT's energy consumption due to additional computation and communication.
We thus, albeit development effort, synchronization complexity and component cost, deploy an STM32F765 MCU as core of our system.
The conceivable \emph{form factor} ranges from a standalone PCB (only containing the measurement system) to a fully integrated solution with both DUT and energy monitoring on a single board.
Although the latter option allows for closer and more precise monitoring, it also limits the system's overall versatility as it is bound to the chosen, single DUT.
The selection of \emph{monitored supply rails} not only dictates the number of ADC channels, but also the coverage of the resulting power trace.
Including relevant I/O components helps to reveal the RTS's temporal behavior w.r.t. communication in addition to its computation/energy footprint.

Combining the two latter design aspects, we thus implement the measurement system on an FMC-compatible board suitable for various pSoC development boards and include relevant I/O interfaces such as dual Gigabit Ethernet and an HDMI input.

\begin{figure}
	\centering
	\includegraphics[width=0.85\linewidth,angle=180]{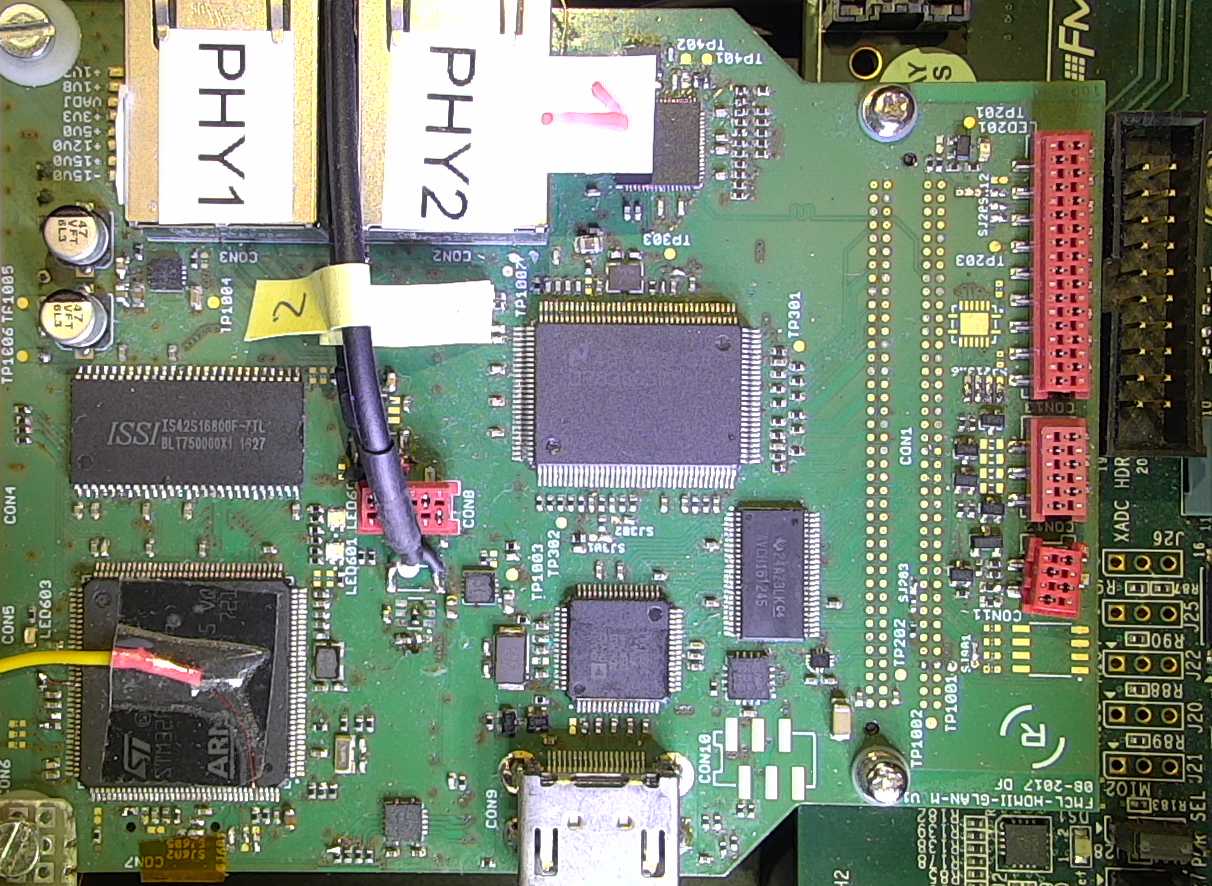}
	\caption{Top View of FMC Prototype with (V/I) Probes routed to ZC702 (left)}
	\label{fig2}
	\vspace{-4.5mm}
\end{figure}

The internals of the measurement system with its interfaces and voltage/current probes to the ZC702 development board are depicted in~Fig.~\ref{fig1}, whilst~Fig.~\ref{fig2} shows the resulting FMC implementation and two (of six) probes connected to the DUT.

\subsection{Measurement System: Hardware and Firmware Details}

Acquisition of voltage/current signals is performed by synchronized, simultaneously sampling 16-bit ADCs (ADS8556) individually capable of up to 630~kSPS.
As all three ADCs and an external SDRAM for extended capture lengths without SD card share the MCU's external data bus, the resulting sampling rate is lowered to around 225~kSPS.
During each iteration, the MCU not only synchronously triggers and sequentially reads all ADCs, but also performs PMBus and SD card transactions as required.
In our current firmware configuration, recordings are automatically started based on a digital trigger line between DUT and MCU.
Even without any event, a pre-trigger buffer continuously captures over 150~ms of analog samples to enable the analysis of I/O and DUT activities before a triggering.
The\linebreak MCU writes the recorded data to the SD card without filesystem for performance/density reasons and adds a precise sub-$\mu$s timestamp per analog/PMBus sample block and trigger event.

The ADCs are configured for an input voltage range of 10~V, which enables direct measurement of all supply rail voltages.
Currents are first converted to usable sense voltage by Current Sense (CS in~Fig.~\ref{fig1}) circuitry consisting of shunt resistor and gain amplifier.
We tap the ZC702's existing INA333-based CS circuitry for DUT rails and integrate 1\% shunts and LMP8640 sense amplifiers for the I/O components on our FMC board.

\subsection{Monitored I/O Interfaces: Gigabit Ethernet and HDMI}
Our FMC prototype also hosts a relatively old TI DP83865 (PHY1) and a Microchip KSZ9031 (PHY2) for Gigabit Ethernet plus an Analog Devices ADV7611 HDMI receiver.
Both PHYs are supported by the Linux kernels used in our lab and require four supply rails for analog, core and I/O each that are monitored separately.
Data are exchanged via 24 signals with two clocks (per PHY) and an additional 17 I/Os including one clock for HDMI whilst management relies on MDIO and $\text{I}^{2}\text{C}$.

\section{Experimental Evaluation}
\noindent
For commissioning and initial evaluation of both measurement system and the two integrated Ethernet PHYs, we implemented a reference design on the Zynq pSoC (Sec.~\ref{subsec1}).
We also give a summary of lessons learned during this process (Sec.~\ref{subsec2}).

\subsection{High-speed Visual Servoing System: Energy and Latencies}
\label{subsec1}
As the desired DUT reference design should not only feature varying load conditions, but also utilize both Ethernet PHYs, we combine a mixed-hardware/software image processing system (acquiring a high-speed video stream from a GigE Vision camera)~\cite{7746761} with an additional Ethernet controller instantiated within the Programmable Logic (PL)~\cite{xilinx-pg051}.
Whilst the former is capable of acquiring video streams received by the hardcore Ethernet controllers of the Zynq's Processor System (PS), the latter is used as a standard Ethernet device by the Linux kernel.

\begin{figure}
	\centering
	\includegraphics[width=\linewidth]{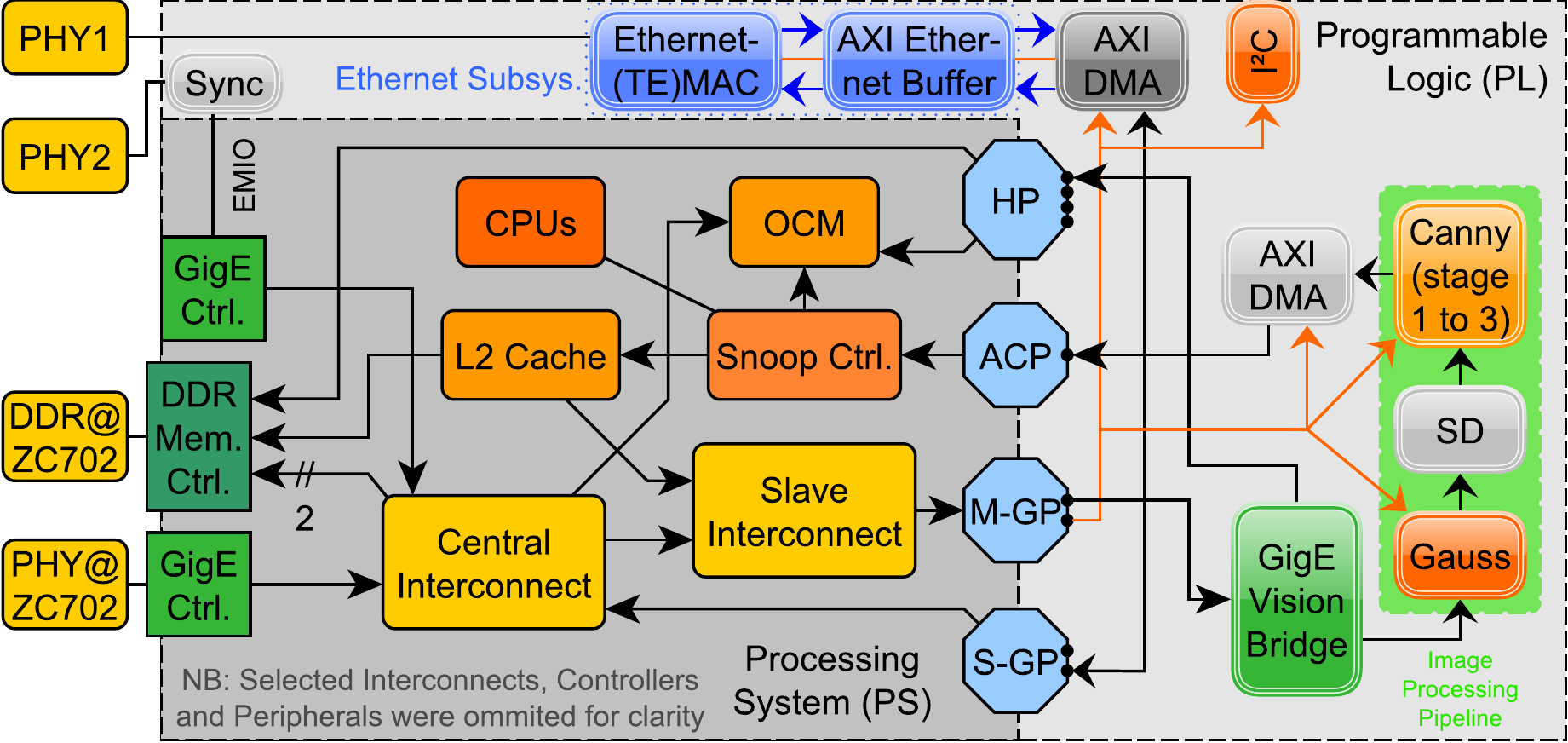}
	\vspace{-4mm}
	\caption{PS Internals \& PL Reference Design for Visual Servoing and 3x GigE}
	\label{fig3}
	\vspace{-4.5mm}
\end{figure}

\begin{figure}
	\centering
	\includegraphics[width=\linewidth]{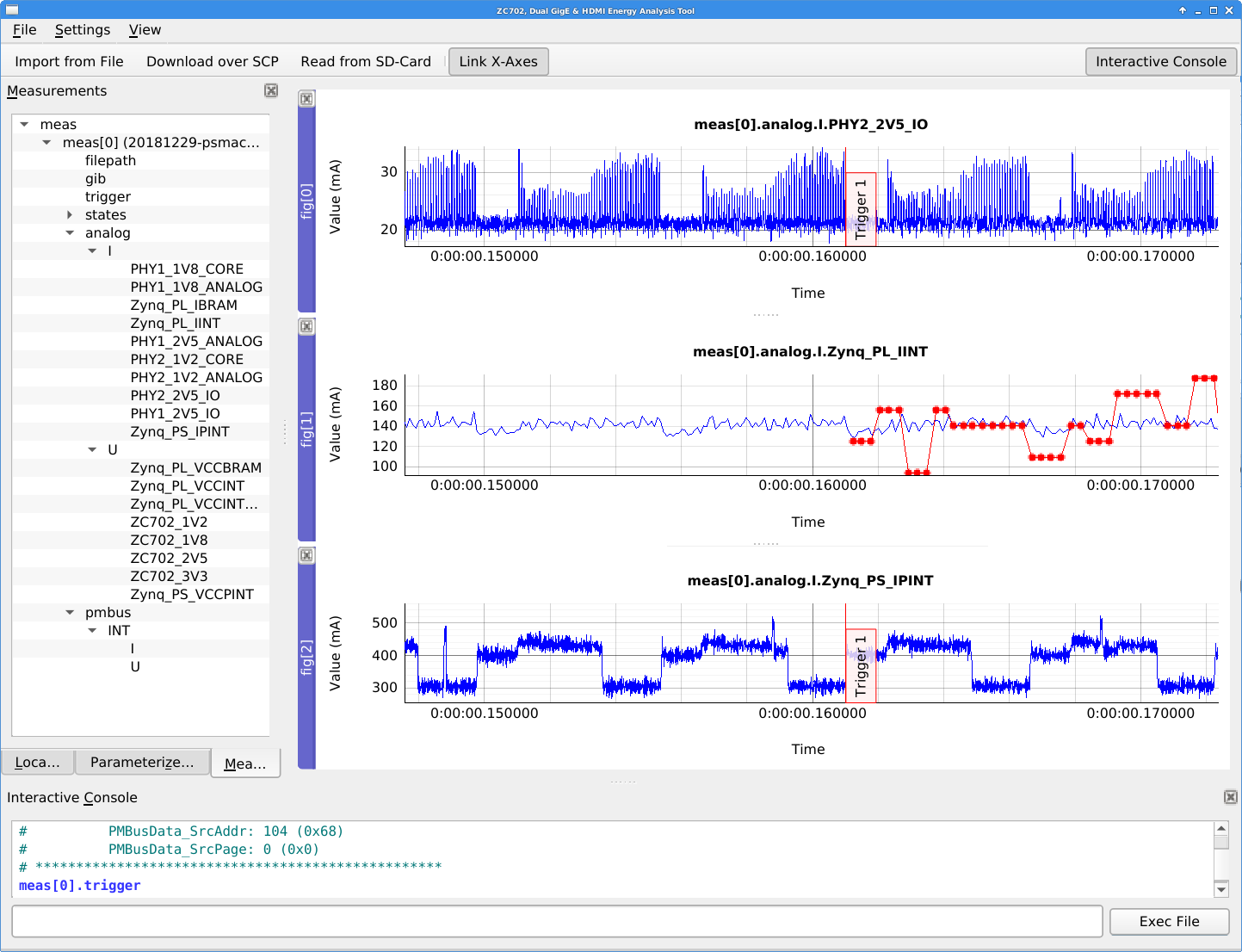}
	\vspace{-4mm}
	\caption{Analysis Tool with Data Tree (left), Plots (right) and Console (bottom): Plots show incoming Ethernet frames (upper) and PL/PS activity (lower two)}
	\label{fig4}
	\vspace{-3.8325mm}
\end{figure}

The resulting design in the PL, selected internals of the PS and external PHYs are shown in~Fig.~\ref{fig3}.
Using our proposed measurement system, we are able to identify individual Ethernet frames sent by the camera (upper plot in~Fig.~\ref{fig4}) and the resulting workload in both PL (middle) and PS (lower).
The trigger was asserted by the PS after successful reception of 1000 camera frames and also marks the point in time where the processing migrates from PHY2 and PL to the PS's CPUs.

The red graph in the PL plot (Fig.~\ref{fig4}, middle) shows the current values from the ZC702's PMBus supply controllers also\linebreak captured by the MCU.
Its low resolution on both axes prohibits detailed PL energy analysis in active and idle system phases.

Our Python-based analysis tool (Fig.~\ref{fig4}) is already capable of various analysis methods in both time and frequency domains and can easily be extended both internally and via its console.

\subsection{FMC Measurement Subsystem: Lessons Learned (so far...)}
\label{subsec2}
Apart from a few quirks (e.g., signal overshoot due to high capacitive load easily rectified by adding series resistors) often found in any first-revision board, there are a couple of lessons w.r.t. the measurement subsystem we have learned so far.

As the system's sampling rate and corresponding low-pass filters' cut-off frequencies exceed 200~kHz, \emph{aliasing artifacts} caused by the switching power supplies operating at 500~kHz become visible.
Active, second-order filters could be used for improved attenuation, but come at the cost of extra amplifiers.

As the measurement system is powered by the FMC connector also serving as a \emph{ground reference} between DUT and FMC, any current drawn also induces a small parasitic voltage offset leading to variable measurement error due to, e.g., SD card activity.
First experiments using a dedicated sense amplifier with ground reference input (INA213) piggybacked on the ZC702, however, show promising results for the sensitive current sense signal of the (surprisingly low-current) BRAM rail.
Although this solution is not optimal due to additional solder joints and resulting voltage offsets caused by the Seebeck effect either, it anyhow enables differential sensing between DUT and ADCs.

Sensing the changes in \emph{I/O component currents} during frame transmission (i.e., from DUT via PHY to the outside) is harder with more modern PHYs that consume a decreasing amount of power during operation.
In our case, we are unable to identify egressing traffic on PHY2, which, in contrast, is clearly visible on PHY1's core rail.
Frame reception, however, can reliably be detected due to the current required on each PHY's VccIO~rail as the PHY drives the parallel data bus from itself to DUT.

\section{Conclusion}
\noindent
In this paper, we proposed a flexible and cost-effective energy monitoring system suitable for various application scenarios.
Relying on a dedicated MCU and an SD card for storage, it not only is operational without DUT intervention and could be integrated for in situ monitoring out in the field, but also supports capture lengths of several hours.
In addition, we proposed to monitor the supply rails of relevant I/O components to gain a holistic view of the system's temporal behavior.
Although the presented FMC-sized implementation of the measurement system itself is still work-in-progress, initial evaluations have already shown its versatility and applicability for both pSoC energy and I/O behavior monitoring of the RTS-under-test.

In contrast to the commonly used power management/monitoring solutions integrated in many pSoC development boards, our solution reaches sampling rates three orders of magnitudes higher and current resolutions as low as a few tens of $\mu$A.
This also enables monitoring of individual Ethernet frames at I/O.

For future work, we intend to increase the measurement system's accuracy via piggybacked gain amplifiers on all DUT channels and continue the evaluation of our RTSs-under-test.
Based on the insights gained w.r.t. both energy and timing, we then will investigate energy management strategies for pSoCs executing mixed-hardware/software real-time applications.

\end{document}